# *Role of Heat Transport in All-Optical Helicity-Independent Magnetization Switching*

V. Raposo[1,2,*], J. Hohlfeld[3], S. Mangin[3], E. Martínez[1,2]

1. *Dpto. Física Aplicada, Universidad de Salamanca, Salamanca, Plaza de los Caídos S/N 37008, Spain.*

2. *Unidad de Excelencia en Luz y Materia Estructuradas (LUMES), Universidad de Salamanca, Salamanca, Spain.*

3. *Institut Jean Lamour, CNRS UMR 7198, Université de Lorraine, Vandoeuvre-lés-Nancy F-54506, France.*

\* Corresponding author: victor@usal.es

Single-shot all-optical helicity independent switching processes are investigated using advanced micromagnetic modeling in a ferrimagnetic thin film embedded in a multilayer stack. Building on recent experimental findings, our multiscale simulations realistically account for heat transport in the stack, focusing on the influence of a metallic copper underlayer with varying thickness. We analyze how this thermal transport affects the final magnetic state of the ferrimagnet as a function of both the laser pulse duration and fluence. Our results reproduce the experimentally observed switching behaviors and elucidate the physical mechanisms that govern the emergence of three distinct final magnetic states. In particular, we demonstrate how these states are critically influenced by the thickness of the underlying copper layer.

# 1. Introduction

Since the first observation of subpicosecond demagnetization in a nickel sample induced by an ultrashort laser pulse—with durations ranging from tens of femtoseconds to several picoseconds—and the subsequent recovery of magnetization [1], extensive experimental [2–18] and theoretical [19–27] efforts have been devoted to understand and control the optical manipulation of magnetic states in ferro- and ferrimagnetic materials. This phenomenon, known as all-optical switching (AOS), enables the reversal of magnetization using light alone, without the need for external magnetic fields.

Most experimental studies have focused on thin films with perpendicular magnetic anisotropy (PMA), where the magnetization at equilibrium points along the out-of-plane ($z$-axis) direction, and different types of AOS have been identified. All-optical helicity-dependent switching (AO-HDS) [2,3] requires a sequence of circularly polarized laser pulses and has been observed in a wide range of materials [4]. AO-HDS has traditionally been attributed to mechanisms such as magnetic circular dichroism (MCD) [12] and/or the inverse Faraday effect (IFE) [13], where, in addition to laser-induced heating, the helicity of the light affects energy absorption and/or generates an effective out-of-plane magnetic field (see [26] for further details). In contrast, all-optical helicity-independent switching (AO-HIS) has been widely observed in Gd-based ferrimagnetic systems, including both amorphous alloys [4–7] and synthetic ferrimagnets [8–11]. In AO-HIS, a single-shot linearly polarized ultrashort laser pulse can locally reverse the magnetization. This process is understood as arising from the distinct demagnetization rates of the two antiferromagnetically coupled sublattices in a ferrimagnet when exposed to a laser pulse of suitable duration and fluence, as described in phase diagrams [28]. The reversal occurs through angular momentum transfer between the sublattices, where one sublattice remains demagnetized while the other begins to recover [17]. More recently, a third mechanism known as all-optical precessional switching (AO-PS) has been identified [29]. In this case, the reorientation of the anisotropy axis from perpendicular to in-plane during laser excitation induces a precession of the magnetization, ultimately resulting in its reversal.

The theoretical description of AO-HIS experimental observations remains, to this day, far from complete. One of the main limitations is that most theoretical studies adopt an atomistic approach [19–24], which is restricted to very small systems with lateral dimensions of only a few nanometers, whereas experimentally studied thin films typically extend over several microns. Furthermore, these atomistic models generally assume uniform laser illumination, making them unable to capture the non-uniform heating effects caused by laser beams with micrometer-scale profiles. Only recently advanced micromagnetic models have been developed to realistically describe AOS processes in both ferromagnetic [25,26] and ferrimagnetic materials [27], considering extended thin films exposed to ultrashort laser pulses with spatially realistic beam profiles. This multiscale framework describes the magnetization dynamics in ferromagnetic (FM) and ferrimagnetic (FiM) systems as they evolve through non-equilibrium thermodynamic states induced by the transient laser excitation. It covers not only the spatial dimensions relevant to actual experiments, but also an extensive range of time scales, from a few femtoseconds to several microseconds. We have previously used this approach to realistically model multi-shot AO-HDS in FM thin films and to investigate the roles of MCD and IFE as the driving mechanisms [26]. More recently, we extended our multiscale micromagnetic solver to study both AO-HIS and AO-HDS in FiM thin films under realistic experimental conditions [27].

In this work, we build on a recent experimental study [7] that investigated the role of a copper (Cu) underlayer beneath a GdFeCo film in AO-HIS processes. The experiment examined the final magnetic state of the GdFeCo alloy as a function of the duration and fluence of a single-shot linearly polarized laser pulse. Using our realistic multiscale modeling approach, we are able to explain how the Cu thickness influences the outcome. In particular, we show that the observed behavior can be naturally interpreted in terms of the transient non-equilibrium thermodynamic states reached in the GdFeCo layer and the dynamics of heat diffusion through the Cu underlayer.

## 2. Micromagnetic model and details

A typical ferrimagnetic (FiM = $RE_x(TM)_{1-x}$) alloy formed by a transition metal (TM ≡ CoFe) and a rare earth (RE ≡ Gd) with a relative concentration of $x = 0.25$ is considered here. As shown in Fig. 1(a), the FiM layer, with a thickness of $t_{FiM} = 10$ nm, is sandwiched between a TaCu top layer and bottom Cu/Pt bilayer, where the thicknesses of the top TaCu and the bottom Pt are fixed ($t_{TaCu} = 10$ nm, $t_{Pt} = 5$ nm). To explore the role of heat diffusion on the laser-induced magnetization dynamics in the stack, the thickness of the Cu layer ($t_{Cu}$) is varied between 5 nm and 50 nm. The stack has a square shape ($\ell \times \ell$) in the $xy$-plane with $\ell$ being the side. Numerically evaluated samples are $\ell \sim 3\, d_0$, where $d_0$ indicates the full width at half maximum of the laser spot, which is centered on the top of the stack.

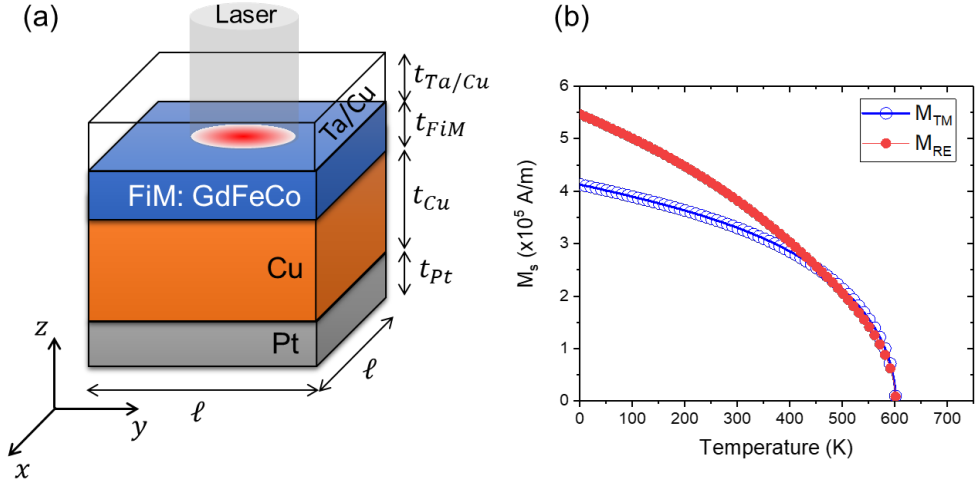

**Figure 1**. (a) Schematic of the multilayer stack explored in the simulations: Pt(5 nm)/Cu($t_{Cu}$)/GdFeCo (10 nm)/TaCu (10 nm)/Pt(5 nm). The copper layer thickness $t_{Cu}$ is varied to investigate its influence on heat transport. (b) Temperature dependence of the saturation magnetization for each sublattice (Gd and FeCo) in the $Gd_{0.25}FeCo_{0.75}$ alloy. These micromagnetic results were obtained under uniform temperature conditions in a small test nano sample and validated against atomistic simulations.

The temporal evolution of the reduced local magnetization $\vec{m}_i(\vec{r},t) = \vec{M}_i(\vec{r},t)/M_{si}(T)$ of each sublattice ($i$: RE ≡ Gd, TM ≡ CoFe) is governed by the stochastic Landau-Lifshitz-Bloch (LLB) equation as coupled to the heat transport obtained from the Two Temperatures Model (2TM) [27]. Note that $\vec{M}_i(\vec{r},t)$ is the local magnetization of sublattice $i$ in units of A/m, and $M_{si}(T)$ is the corresponding spontaneous magnetization at temperature $T$. The LLB eq. reads as

$$\frac{\partial \vec{m}_i}{\partial t} = -\gamma'_{0i}(\vec{m}_i \times \vec{H}_i) - \frac{\gamma'_{0i}\alpha_i^\perp}{m_i^2}\vec{m}_i \times [\vec{m}_i \times (\vec{H}_i + \vec{\xi}_i^\perp)] + \\ + \frac{\gamma'_{0i}\alpha_i^\parallel}{m_i^2}(\vec{m}_i \cdot \vec{H}_i)\vec{m}_i + \vec{\xi}_i^\parallel + \vec{\tau}_i^{NE} \qquad (1)$$

where $\gamma'_{0i} = \gamma_{0i}/(1 + \lambda_i^2)$ is the reduced gyromagnetic ratio, which is defined via the coupling parameter $\lambda_i$ of sublattice $i$ to the heat bath. $\alpha_i^\parallel$ and $\alpha_i^\perp$ are the longitudinal and perpendicular damping parameters. $\vec{H}_i = \vec{H}_i(\vec{r},t)$ is the local effective field at location $\vec{r}$ acting on sublattice magnetic moment $i$. It includes exchange, magnetic anisotropy and dipolar interactions. $\vec{\xi}_i^\parallel$ and $\vec{\xi}_i^\perp$ are the longitudinal and perpendicular stochastic thermal fields. Finally, $\vec{\tau}_i^{NE}$ represents the non-equilibrium torque that accounts for the angular momentum transfer between lattices in non-equilibrium states. The magnitude of $\vec{m}_i(\vec{r},t)$ depends on the electron temperature ($T_e(\vec{r},t)$), which is obtained from the 2TM as described below. All details of this extended micromagnetic model can be found in a previous publication by us [27].

Typical material parameters for $Gd_{0.25}(FeCo)_{0.75}$ were adopted for preliminary atomistic simulations: $J_{0,TM-TM} = 2.59 \times 10^{-21}$ J (ferromagnetic exchange energy between two TM atomic magnetic moments), $J_{0,RE-RE} = 1.35 \times 10^{-21}$ J (ferromagnetic exchange energy between two TM atomic magnetic moments), $J_{0,TM-RE} = -1.13 \times 10^{-21}$ J (antiferromagnetic exchange energy between TM and RE magnetic moments), $d_u = 8.045 \times 10^{-23}$ J (uniaxial PMA energy), $\lambda_{TM} = \lambda_{RE} = 0.02$. The atomic magnetic moments are $\mu_{TM} = 1.92\mu_B$ and $\mu_{RE} = 7.63\mu_B$, where $\mu_B$ is the Borh magneton. The lattice constant is $a = 0.352$ nm. All these symbols were defined in [27]. Micromagnetic parameters needed to solve the LLB Eq. (1) can be obtained from atomistic inputs as follows [20]. The spontaneous magnetization at zero temperature of each sublattice is $M_{si}(0) = \mu_i x_i/(p_f a^3)$, where $x_i$ is the concentration of sublattice $i,j$: RE, TM, ($x_j = 1 - x_i$), and $p_f$ is a packing factor which depends on the crystalline structure ($p_f = 0.74$ for fcc, [20]): $M_{S,TM}(0) = 0.412$ MA/m, $M_{S,RE}(0) = 0.546$ MA/m. The continuous exchange stiffness constants [22] are $A_{ex,i} = n\epsilon x_i^2 J_i/(2a)$ with $n = 2$ being the number of atoms per unit cell for fcc and $\epsilon = 0.79$ the spin wave mean field correction value also for fcc: $A_{ex,TM} = 3.27$ pJ/m, $A_{ex,RE} = 0.189$ pJ/m. The perpendicular magnetic anisotropy parameter is calculated as $K_i = d_u x_i/(p_f a^3)$, see [20], so $K_{u,TM} = 1.87$ MJ/m³, $K_{u,RE} = 0.62$ MJ/m³. The Landé factors are $g_{TM} = 2.1$ and $g_{RE} = 2.0$. Gilbert damping parameters are $\alpha_{TM} = 0.01$ and $\alpha_{RE} = 0.005$. For the presented results we take $\lambda_{ex} = 0.2$ for the exchange relaxation rate (see details in [27]). Fig. 1(b) shows temperature evolution of the saturation magnetization of the two sublattices as obtained from the LLB eq. (1) for a small FiM nanodot under uniform temperature. These micromagnetic results are in quantitative agreement with corresponding results obtained from atomistic simulations in such a small nanosample, and serve as proof confirming the validity of the LLB eq. based micromagnetic framework (see [27] for further details).

Returning to realistic microscale samples, we consider an initial uniform magnetic state in the FiM with the magnetic moments of the two sublattices aligned antiparallel to each other along the easy axis (z-axis). Upon application of a laser pulse, the irradiated stack absorbs energy and excites the magnetization dynamics. The laser spot is assumed to have a spatial Gaussian profile ($\eta(r)$), with $r_0$ being the radius spot ($d_0 = 2r_0 = 8.1$ µm is the full width at half maximum FWHM of the laser spot). Its temporal profile ($\xi(t)$) is also Gaussian, with $\tau$ representing the pulse duration (FWHM). The absorbed power density is expressed as $P(\vec{r}, t) = P_0 \eta(r) \xi(t)$ where $\eta(r) = \exp\left[-\frac{4 \ln(2) r^2}{(2r_0)^2}\right]$ is the spatial profile, with $r = \sqrt{x^2 + y^2}$ being the distance from the center of the laser spot, and $\xi(t) = \exp\left[-\frac{4 \ln(2)(t - t_0)^2}{\tau^2}\right]$ is the temporal profile. $P_0$ is the maximum power density of the laser pulse reached at $t = t_0$ at the center of the laser spot located at $(x_c, y_c) = (0,0)$. The fluence $F$ indicated in the text is proportional to $P_0$, but the exact relation depends on the effective penetration depth ($\delta_s$) and the absorption coefficient of the sample ($A$): $F \approx A P_0 \tau \delta_S$. In the preset study we considered $\delta_s = 20$ nm, $A = 0.5$.

Laser pulse heats the multilayer stack (TaCu/GdFeCo/Cu/Pt) from the top, and consequently, the system is transiently dragged into a nonequilibrium thermodynamic state, where the GdFeCo magnetization changes according to the electron temperature dynamics. The temporal evolution of the temperature is described by the two temperatures model (2TM) [27] in terms of two subsystems: the electron ($T_e = T_e(\vec{r}, t)$) and the lattice ($T_l = T_l(\vec{r}, t)$),

$$C_e \frac{\partial T_e}{\partial t} = k_e \nabla^2 T_e - g_{el}(T_e - T_l) + P(r, t) - \frac{T_e - T_R}{\tau_D} \quad (2)$$

$$C_l \frac{\partial T_l}{\partial t} = -g_{el}(T_l - T_e) \quad (3)$$

where $C_e$ and $C_l$ denote the specific heat of electrons and lattice subsystems, respectively. $k_e$ is the electronic thermal conductivity. $g_{el}$ is a coupling parameter between the electron and lattice subsystems, and $\tau_D$ is the characteristic heat diffusion time to the substrate and the surrounding media. $T_R = 300$ K is room temperature. Eqs. (2) and (3) are numerically solved for each layer of the stack with their corresponding thermal parameters ($C_e$, $C_l$, $k_e$ and $g_{el}$). $C_e = C_e(T_e)$ is linear with the electron's temperature $T_e$ as $C_e(T_e) = \gamma_e T_e$ where $\gamma_e = 600$ J/(m$^3 \cdot$ K$^2$) is the electronic heat capacity. Due to the interfacial thermal resistance at the TaCu/FiM and FiM/Cu interfaces, a reduced thermal conductivity of 15 W/(K $\cdot$ m) is imposed. We also assume that the laser power density ($P(r, t)$) is only partially absorbed from the top Ta/Cu and the FiM layers. The Newton-like term (last term in Eq. (2)) accounts for thermal relaxation towards room temperature. The following layer-dependent thermal parameters were assumed. Thermal conductivities ($k_e$) are: 80 W/ (K $\cdot$ m) for the Top TaCu layer, $k_e = 91$ W/ (K $\cdot$ m) for the FiM layer, and $k_e = 300$ W/ (K $\cdot$ m) for the bottom Cu and Pt layers. The lattice specific heat ($C_l$) are: $3.44 \times 10^6$ J/ (K $\cdot$ m$^3$) for both the top and bottom TaCu, Cu and Pt layers, and $3.8 \times 10^6 \text{J}/(\text{K} \cdot \text{m}^3)$ in the FiM. The electron-lattice coupling parameter is $g_{el} = 7 \times 10^{17}$ W/m³ for all layers in the stack.

In the presented micromagnetic framework the laser-induced magnetization dynamics is evaluated by numerically solving the LLB Eq. (1) in the GdFeCo layer as coupled to the 2TM Eqs. (2)-(3) (in the full TaCu/FiM/Cu/Pt stack) using a Heun algorithm with an adaptative time step of $\Delta t = 1$ fs for $0 \leq t \leq 10\ t_0$, and of $\Delta t = 50$ fs for $t > 10\ t_0$, with $t_0$ being the time at which the laser pulse reaches its maximum power. The full TaCu/FiM/Cu/Pt stack, with dimensions $\ell \times \ell \times (t_{TaCu} + t_{FiM} + t_{Cu} + t_{Pt})$, is discretized using a finite differences scheme using computational cells of volume $\Delta V = \Delta x \Delta y \Delta z$ with $\Delta x = \Delta y = 3$ nm and $\Delta z = 5$ nm. The micromagnetic model solving the coupled magneto-thermal problem in the TaCu/FiM/Cu/Pt stack was implemented on a home-made CUDA-based code as described in [27], and run on a RTX3090 GPU. It was also checked in several tests that decreasing the cell size to $\Delta x = 1$ nm does not change the presented results.

## 3. Micromagnetic results

The numerically obtained phase diagrams in Fig. 2(a) show the three distinct final magnetic states as a function of laser pulse duration ($\tau$) and fluence ($F$), for three different copper underlayer thicknesses ($t_{Cu}$). In qualitative agreement with experimental observations [7], our micromagnetic simulations reveal the same three outcomes: no switching (NS), switching (SW), and multidomain (MD) final states. Representative examples of these final magnetic configurations are presented in Fig. 2(b) for three representative ($\tau$, $F$) combinations, as indicated in Fig. 2(a)–(b) and the caption.

We find that the critical pulse duration required to achieve switching increases linearly with fluence, while the NS→SW transition is largely unaffected by the Cu underlayer thickness. In contrast, the transition from SW to MD occurs for any value of $\tau$ when the fluence exceeds a certain threshold, as illustrated by the vertical red dotted line for $t_{Cu} = 5$ nm. This SW→MD boundary shifts toward higher fluence values as the Cu layer becomes thicker, which is also in good qualitative agreement with the experimental results reported in Ref. [7].

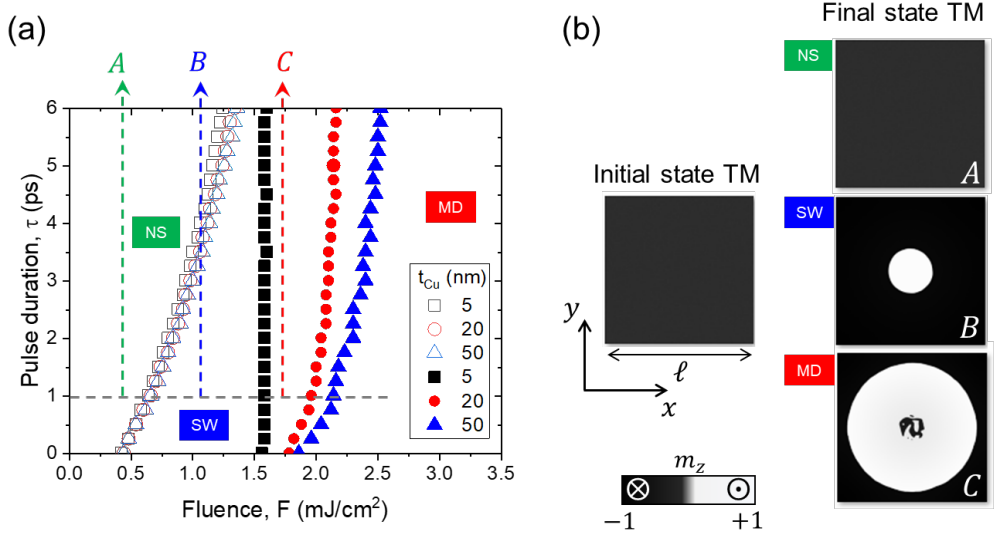

**Figure 2**. (a) Phase diagrams showing the final magnetic state of the GdFeCo ferrimagnetic layer as a function of laser pulse duration ($\tau$) and fluence ($F$), for three different copper underlayer thicknesses ($t_{Cu}$). Three representative points labeled $A$, $B$, and $C$ correspond to different combinations of ($\tau$, $F$) values: $\tau = 1$ ps with $F = 0.4$, $1.1$, and $1.7$ mJ/cm$^2$, respectively, illustrating the no switching (NS), switching (SW), and multidomain (MD) regimes. (b) Snapshots of the out-of-plane component of the transition metal (TM) magnetization, ($m_{z,TM}(\vec{r},t)$), at the initial state ($t = 0$) and final state ($t = 100$ ns), corresponding to points $A$, $B$, and $C$ in panel (a).

In order to provide further insight into the origin of the results presented in Fig. 2, we have analyzed the temporal evolution of the electron temperature ($T_e(0,t)$) in the center of the GdFeCo sample upon application of a fixed laser pulse ($\tau = 1$ ps, $F = 0.76$ mJ/cm$^2$) when it is placed on top a Cu underlayer with different thicknesses. The results, obtained from the 2TM (Eq. (2)-(3)), are shown in Fig. 3 which plots the laser power at the center of the laser spot (left vertical axis) and the corresponding $T_e(0,t)$ (right vertical axis). Note that the horizontal time axis is shown on a logarithmic scale to evidence the wide range of different time scales taking place in these AOS processes. Three transient regimes are clearly shown. During the application laser pulse, the temperature increases abruptly from room temperature ($T_R = 300$ K) to a peak value of $T_e \sim 950$ K, which is significantly larger than the Curie temperature of the FiM ($T_C = 600\ K$, see Fig. 1(b)). Still within regime *I*, the local $T_e$ decreases very fast once the laser pulse is turned off. Indeed, $T_e$ reaches a value below the Curie threshold within the first picoseconds after the pulse (see Fig. 3 around $t = 10$ ps). Note that this regime *I* is essentially independent of the Cu thickness, as the three cases shown in Fig. 3 show the same temporal dynamics of the electron temperature for $t \lesssim 7$ ps. Therefore, Cu underlayer is still not affecting the temperature in FiM in regime *I*. However, Cu becomes relevant for longer times after the laser pulse. A significantly slower temperature dynamics is observed in regime *II* (10 ps $\lesssim t \lesssim$ 1 ns), where the heat transport in the Cu layer starts to take place. As expected, the thicker the Cu layer, the smaller the electron temperature of the FiM in this intermediate regime *II*. Finally, for even longer times, the $T_e$ relaxes toward to room temperature in the regime *III*, which occurs in the time range of 10 ps ≲ t ≲ 1 ns. As it will be clear later on, this full thermal dynamic is essential to understand the final state achieved by the GdFeCo as shown in previous Fig. 2(a).

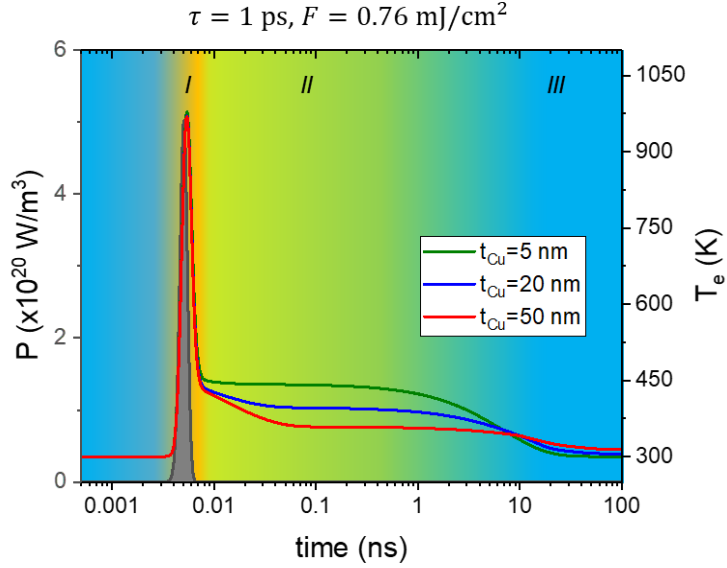

**Figure 3**. Temporal evolution of the laser power at the center of the laser spot (left vertical axis, $P$) and the electron temperature at the center of the FiM (right vertical axis, $T_e(0,t) \equiv T_e$) for three different Cu thickness underlayer. The laser pulse has fixed duration of $\tau = 1$ ps and fixed fluence of $F = 0.76$ mJ/cm$^2$. The labels in roman numbers ($I, II, III$) indicate the three transient regimes taking place during the full process. Background color is added to visually assist on the description.

In top graphs of Fig. 4 we plot the temporal evolution of the laser power at the center of the laser spot ($P \equiv P(0,t)$) and the electron temperature in the center of the FiM ($T_e \equiv T_e(0,t)$), just below the center of the laser beam for the representative cases $A$, $B$ and $C$ marked in Fig. 2(a). These cases correspond to a laser pulse of $\tau = 1$ ps and three different values of the fluence ($F$) as indicated at the top of each column in Fig. 4. The vertical and the horizontal lines shown in these graphs indicate the time ($t^*$) at which the temperature of the FiM reaches the Curie threshold ($T_C = 600$ K) during the cooling down after effect. These times ($t^*$) mark the instant at which the magnetic order starts its recovery after the strong temperature values reached due to the laser pulse, and they are relevant to understand the final state of the FiM. The corresponding temporal evolution of the averaged out-of-plane component of the magnetization of the each sublattice of the FiM ($m_{z,TM}$ and $m_{z,RE}$) are shown in bottom graphs of Fig. 4 for such $A$, $B$ and $C$ cases. Such values were obtained by averaging the magnetization of each sublattice over a centered circular area with a diameter ($d$) smaller than the full width at half maximum of the laser beam ($d = 0.25\, d_0$).

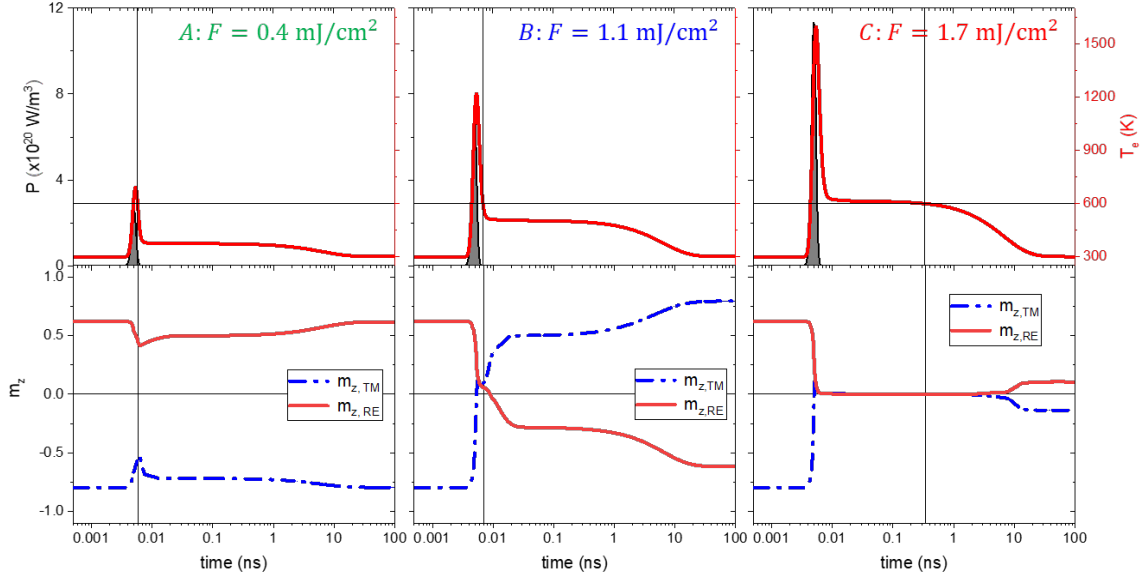

**Figure 4.** Top graphs show the temporal evolution of the laser power at the center of the laser spot ($P \equiv P(0, t)$, left vertical scale) and the electron temperature in the center of the FiM ($T_e \equiv T_e(0, t)$, right vertical scale), just below the center of the laser beam for the representative cases $A$ (left column), $B$ (center column) and $C$ (right column) already marked in Fig. 2(a). These cases correspond to a laser pulse of $\tau = 1$ ps and three different fluence values ($F$) as indicated on the corresponding top graph. Bottom graphs show the corresponding temporal evolution of the average out-of-plane component of the magnetization of the each sublattice of the FiM ($m_{z,TM}$ and $m_{z,RE}$).

Case $A$ is a representative example of non-switching (NS) behavior. Its dynamics is shown in left panel of Fig. 4, which shows that the temperature overcomes the Curie threshold during the laser pulse, and it recovers that value at $t_A^* \simeq 5.8$ ps (see vertical line in left graphs). Although a small variation of the average out-of-plane magnetization is evidenced in left-bottom graph, neither of the two sublattices becomes demagnetized nor can reverse its initial sense in the time scale around the laser pulse (1 ps − 10 ps). Therefore, the fluence ($F = 0.4$ mJ/cm$^2$) is not enough to promote the final switching, so the FiM goes back to its initial state once the full thermal relaxation ends. The switching of the FiM (SW) is achieved for intermediate values of fluence, as exemplary depicted by case $B$ ($F = 1.1$ mJ/cm$^2$) presented in the central graphs of Fig. 4. Now the maximum temperature is larger than in case $A$, and the ferrimagnetic order is recovered later $t_B^* \simeq 7$ ps. In that moment the TM magnetization has already reversed its initial sense (see dashed blue line in bottom central graph), whereas the RE is only slightly demagnetized without reversing its initial sense. As time elapses from $t > t_B^* \simeq 7$ ps, the temperature decreases below the Curie threshold ($T_C = 600$ K), and the FiM starts to recover the magnetic order being dominated by the TM sublattice because its saturation magnetization is higher than the one by the RE in this temperature regime (see Fig. 1(b) for $450 \text{ K} \lesssim T_e \lesssim 600 \text{ K}$). After this short transient ferromagnetic-like state, and as temperature decreases below $T_C$, the ferrimagnetic order becomes to be recovered, and the already reversed TM magnetization drives the reversal of the RE due to the antiferromagnetic exchange. Both sublattices depict an intermediate slow dynamic linked to regime *II* described in Fig. 3, where the heat transport over the full stack is taking place. The final switching state is achieved at longer times (~50 ns), where the full system has recovered its thermal equilibrium at room temperature.

The third behavior is the final multidomain case (MD), which occurs for much higher fluences, as shown in case *C* at the right panel of Fig. 4 ($F = 1.7 \text{ mJ/cm}^2$). In this case, and differently from the case *B* (SW), both sublattices become fully demagnetized because of the laser pulse ($m_{z,TM} \approx m_{z,RE} \approx 0$), and they remain in such demagnetized state without any magnetic order until the Curie temperature is recovered at $t_C^* \simeq 340$ ps (see vertical lines in the right panel of Fig. 4, $t_C^* \simeq 340$ ps). Note that $t_C^*$ is more than two order of magnitude larger than $t_B^*$ and $t_A^*$. Thermal relaxation to the room temperature occurs from time $t \gtrsim 1$ ns. During this relaxation, the fully demagnetized systems recover the ferrimagnetic order by adopting a random multidomain pattern which become energetically stable from $t \sim 10$ ns to $t \sim 100$ ns. The finite final values of $m_{z,TM}$ and $m_{z,RE}$ are representative of such stable MD pattern.

## 4. Concluding remarks

The representative cases *A*, *B*, and *C*, along with the corresponding magneto-thermal dynamics presented in Fig. 3 and Fig. 4, provide a comprehensive understanding of the AO-HIS phase diagram shown in Fig. 2 and its dependence on the thickness of the copper underlayer ($t_{Cu}$) Importantly, our micromagnetic modeling successfully reproduces the experimentally observed phase diagram [7] and offers clear insight into the physical mechanisms that govern the transitions between the different magnetic states. The transition from no switching (NS) to switching (SW) is found to be essentially independent of $t_{Cu}$ as it occurs when the dominant sublattice—just below the Curie temperature—reverses its magnetization on a timescale of approximately $2\tau$–$3\tau$. In this regime, heat transport into the Cu underlayer plays a negligible role, and the switching is driven solely by the local temperature dynamics within the GdFeCu layer. In contrast, the transition from SW to the multidomain (MD) state is strongly influenced by heat transport across the multilayer stack. Above a certain fluence threshold, the FiM becomes fully demagnetized ($m_{z,TM} \approx m_{z,RE} \approx 0$) shortly after the laser pulse, while the local temperature remains above the Curie point for an extended time. The recovery from this demagnetized state—and whether the system reorders into a single-domain or multidomain configuration—is governed by the rate of heat diffusion. A thicker Cu underlayer accelerates the cooling process of the FiM, allowing earlier re-establishment of ferrimagnetic order. Consequently, the critical SW→MD boundary shifts toward higher fluence values as $t_{Cu}$ increases, in good agreement with the experimental findings. Our multiscale micromagnetic modeling not only captures the experimentally observed AO-HIS regimes with high fidelity but also clarifies the underlying magneto-thermal dynamics responsible for each final state. These insights are expected to be valuable for the future design of ultrafast optically controlled spintronic devices operating on sub-picosecond timescales.

## Acknowledgements

The work by V.R. and E.M. was partially supported by project PID2023-150853NB-C31 funded by MICIU/AEI /10.13039/501100011033 and by FEDER, UE. This work was supported by the ANR through the France 2030 government grants EMCOM (ANR-22-PEEL-0009), PEPR SPIN (ANR-22-EXSP 0002) and PEPR SPIN – SPINMAT ANR-22-EXSP-0007 and ANR-23-CE30-0047 SLAM, the Institute Carnot ICEEL, the Région Grand Est, the Metropole Grand Nancy for the project "OPTIMAG" and FASTNESS, the interdisciplinary project LUE "MAT-PULSE", part of the French PIA project "Lorraine Université d'Excellence" reference ANR-15-IDEX-04-LUE. This article is based upon work from COST Action CA23136 CHIROMAG, supported by COST (European Cooperation in Science and Technology).

## Data Availability

The data that support the findings of this study are available from the corresponding author, [VR], upon reasonable request.